\documentclass[useAMS, usenatbib]{mn2e}
\usepackage[dvips]{graphicx}
\usepackage{amssymb}
\usepackage{multirow}

\input{epsf}

\newcommand{\kms}{\thinspace km \thinspace s$^{-1}$}

\newcommand{\ha}{\mbox{H$\alpha$}}

\newcommand{\lya}{\mbox{Ly$\alpha$}}
\newcommand{\OIII}{\mbox{[O\thinspace{\sc iii}]}}
\newcommand{\kmsmpc}{\thinspace km\thinspace s$^{-1}$\thinspace Mpc$^{-1}$}

\title{Lyman-break galaxies: high mass or low?}
\author[S. J. Weatherley \& S. J. Warren]
       {Stephen~J.~Weatherley\thanks{Email : stephen.weatherley@imperial.ac.uk} and Stephen~J.~Warren\\
        Astrophysics Group, Blackett Laboratory, Imperial College London, Prince Consort Road, London SW7 2BW, UK}
	
\date{Accepted 0000 January 00.
      Received 0000 January 00;
      in original form 0000 January 00}

\pagerange{\pageref{firstpage}--\pageref{lastpage}}
\pubyear{2003}
\begin{document}
\maketitle
\label{firstpage}


\begin{abstract}

We reassess the hypothesis that Lyman-break galaxies (LBGs) at redshifts
$z\sim3$ mark the centres of the most massive dark matter haloes at
that epoch. First we reanalyse the kinematic measurements of Pettini
et al., and of Erb et al., of the rest-frame optical emission lines of
LBGs. We compare the distribution of the ratio of the rotation
velocity to the central line width, against the expected distribution
for galaxies with random inclination angles, modelled as singular
isothermal spheres. The model fits the data well.
On this basis we argue that the central line width provides a
predictor of the circular velocity at a radius of several
kpc. Assembling a larger sample of LBGs with measured line widths, we
compare these results against the theoretical $\Lambda$CDM rotation
curves of Mo, Mao \& White, under the hypothesis that LBGs mark the
centres of the most massive dark matter halos. We find that the
circular velocities are over-predicted by a substantial factor, which
we estimate conservatively as $1.8\pm0.4$. This indicates that the
model is probably incorrect. The model of LBGs as relatively
low-mass starburst systems, of Somerville, Primack, and Faber (2001),
provides a good fit to the data.

\end{abstract}




\section{Introduction}

The rotation curves of galaxies have played, and continue to play, an
important role in cosmology. The flat rotation curves of spiral
galaxies at large radii (Babcock, 1939; Rubin, Thonnard \& Ford, 1978;
van Albada et al., 1985) were central to the construction of the
modern paradigm that non-baryonic collisionless cold dark matter (CDM)
dominates the matter density of the Universe. Within this paradigm,
for specified cosmological parameters and a spectrum of initial
density perturbations, and ignoring the baryons, the growth of
structure in the dark matter can be modelled accurately (Lacey \&
Cole, 1993; Navarro, Frenk \& White, 1997), providing predictions of
the mass spectrum of dark matter halos at any redshift, and their
density profiles. Recently, the rotation curves of low surface
brightness galaxies have been used to test the predicted density
profiles. Indeed, it has been argued that the inferred constant density cores
are inconsistent with the cuspy
density profiles predicted (de Blok, Bosma \& McGaugh, 2003), perhaps
indicating that a modification of the CDM paradigm is
required. Alternatively the discrepancy may be an indication of the
lack of understanding of how the density profile in the halo core is
modified by the baryons, firstly in the collapse phase, as well as in
any subsequent phase of ejection of the baryons in a wind following
star formation. Therefore there is a tie between progress in
understanding the nature of the dark matter, and progress in
understanding the assembly of the baryonic component of galaxies.

A similar tie exists in the study of galaxies at high redshift. The
development of a successful method for identifying galaxies at
$z\sim3$, the Lyman-break technique (Steidel \& Hamilton, 1993), and
the measurement of the properties of the galaxies discovered (Steidel
et al., 1996; 1998), allow further tests of the CDM paradigm. A simple
hypothesis is that these Lyman-break galaxies (LBGs) mark the centres
of the most massive dark matter halos. The benefit of this hypothesis
is its simplicity, allowing detailed predictions. For example, by integrating
the halo mass function down to the mass limit that produces a number
density matching the number of LBGs, the distribution of halo masses
is known, and the clustering strength can be predicted. Satisfactory agreement
is found between the predicted clustering amplitude, and the measured
value (Giavalisco et al., 1998; Adelberger et al., 1998, Mo, Mao \&
White, 1999).

An alternative view, promoted by Somerville, Primack, and
Faber (2001, hereafter SPF) is that LBGs are starburst galaxies triggered by
collisions. Since the starburst phase is short, under this hypothesis
more halos contain a LBG at some time, and so the halo masses
associated with LBGs are smaller. Kolatt et al. (1999), and Wechsler et al. (2001) show that this
picture also provides a satisfactory explanation of the clustering
properties of LBGs. Although many of the ingredients of this picture
are uncertain, since galaxy collisions are in the nature of
hierarchical galaxy formation, it would be surprising if there were not
some truth in it.

To learn more about how the baryons in galaxies are assembled, and the
nature of the dark matter, we need to discriminate between these two
hypotheses. Clearly measurement of the masses of LBGs is the key to
progress on these issues. Most of the useful kinematic information on
LBGs comes from the measured line widths of rest-frame optical
emission lines. However, because the stellar light in LBGs is very compact, with
half-light radii of typically 0.2 arcsec (Marleau \& Simard, 1998),
these measurements sample the galaxy mass over the central few kpc
only, where the mass is dominated by the baryons, and therefore where
the circular velocity cannot be predicted reliably. For these reasons
the measurement of a handful of rotation velocities at radii of nearly
1 arcsec, by Pettini et al. (2001), and Erb et al. (2003), are
particularly useful, allowing direct comparison against the theoretical halo 
rotation curves, rather than the uncertain predicted stellar central velocity dispersions.

In this paper we reanalyse the kinematic data on LBGs, paying
particular attention to those galaxies for which rotation velocities
have been measured. We compare the results of this analysis against
the predictions of Mo et al., (1999, hereafter MMW) of the halo rotation
curves of LBGs under the hypothesis that LBGs mark the centres of the
most massive dark matter halos. In Section 2 we assemble the kinematic
data. In Section 3 we analyse the data on rotation velocities. We
compare the distribution of the ratio of the rotation velocity to the
central line width, against the expected distribution for galaxies
with random inclination angles, modelled as singular isothermal
spheres. The model fits the data well. On this
basis we argue that the distribution of central line widths provides a
reliable predictor of the distribution of circular velocities at radii
of several kpc. We use this to infer the distribution of circular
velocities for a larger sample of galaxies, for which only central
line widths have been measured. In Section 4 we compare these circular
velocities against the predictions of MMW, finding that the measured
circular velocities are substantially smaller than the predicted
values. Consideration of possible biases only reinforces this
conclusion. We then compare our results against the collisional starburst model
predictions of SPF, and find good
agreement. Throughout, we assume a standard, flat $\Lambda$CDM
cosmology with $\Omega_\Lambda = 0.7$ and H$_0 = 70$\kmsmpc. The 
same cosmology was used for the two models we compare against. For
this cosmology an angle of 1 arcsec corresponds to the physical size
8.4 kpc at $z=2$, and 7.7 kpc at $z=3$.


\section{LBG kinematic data}

In this section we assemble suitable kinematic data on LBGs for
analysis. The rest-frame optical emission lines are the most useful
for quantifying gravitational motions in LBGs. The \lya\ emission line
cannot be used because it is broadened by resonant scattering, while
the strong interstellar absorption lines visible in rest-frame UV
spectra are probably broadened by non-gravitational motions and, in
any case, are saturated. Any genuine stellar absorption lines are too
weak to be useful.

Most of the kinematic information on LBGs is restricted to the
measurement of the line widths. This is because, even in good seeing
($\sim0.5$ arcsec), from the ground the light profiles of most LBGs
are spatially unresolved. There is still a relative paucity of these
kinematic measures, largely due the difficulty in detecting rest-frame
optical emission which has been redshifted into the near-infrared.
Therefore the strongest lines, \OIII\ and \ha, are the most
useful. The two largest samples of near-infrared spectra of LBGs are
those of Pettini et al. (2001, hereafter P01), containing kinematic
data on 16 galaxies, and of Erb et al. (2003, hereafter E03), also
containing kinematic data on 16 galaxies. In Table 1 we summarise the relevant
data on these 32 LBGs. Col. (1) provides the galaxy name, col. (2) the
galaxy redshift, col. (3) the $\cal R$ apparent magnitude, and
col. (4) the emission-line velocity width, characterised by the
standard deviation $\sigma$ of a Gaussian fit, and its uncertainty (if
provided in the original paper). In a few cases only upper limits are
listed, as the lines were unresolved at the resolution of the
observations. The emission line measured is listed in col. (5), and
col. (6) provides the reference.

In a small number of cases the emission lines are spatially resolved,
and show evidence of rotation. As argued above, these galaxies provide
particularly valuable kinematic information. As emphasised by E03 the
measurement of rotation in LBGs from the ground is difficult, because
usually the galaxy half-light radius is smaller than the seeing. The
two-dimensional structure of the emission line in the spectrum is
therefore dominated by the kinematics near the centre of the
galaxy. The effect of this is to smooth out the transition from $+$ve
to $-$ve velocity in the rotation curve. For this reason we disregard
the actual shape of the LBG rotation curves
(e.g. Fig. 5, E03), and use only the quoted rotation velocity for each
galaxy, taken as half the velocity difference between the two extreme
points measured. Systematic biases in the measured rotation velocities
are discussed in Section 4. Whether or not the rotation curve continues to
rise after the last point does not concern us, since we will compare
against predictions at the measured radii.

Details of the kinematic measures for the subset of eight galaxies
with rotation velocities are provided in Table 2. In each case the
authors provide both the central line width $\sigma$, and the rotation
velocity $V_r$. In Table 2 the first four columns are the same as the
first four columns of Table 1. Col. (5) lists the rotation velocity
$V_r$, and col. (6) gives the ratio of the rotation velocity $V_r$ to
the central line width $\sigma$. In this paper we use the term
rotation velocity $V_r$ to refer to the measured projection of the
circular velocity $V_c$. For an edge-on galaxy $V_r=V_c$. For a galaxy
with angle of inclination $i<90^{\circ}$, the ratio $V_r/V_c$ depends
on $i$, as well as the alignment of the slit relative to the major
axis. The angular displacement from the galaxy centre at which $V_r$
is measured, $\theta$, is listed in col. (7), and the corresponding
projected distance, $r$, is listed in col. (8)\footnote{In this paper
we use the symbol $r$ for projected radius, and $R$ for deprojected
radius.}. The last three columns list the slit width of the
spectroscopic measurement (col. 9), the emission line measured
(col. 10), and the reference (col. 11).

An important point to note is that, with one exception, the usual
procedure of aligning the slit with the galaxy major axis was not
followed for these galaxies. In most cases this was because only
ground-based imaging was available at the time of the spectroscopic
observations, and therefore there was no useful information on the
galaxy orientations. Therefore, in the analysis of rotation velocities
(Section 3), we assume a random angle of the slit relative to the
galaxy major axis. The one exception is the galaxy SSA22a-MD41. This
galaxy is included in Table 2 for completeness, but is excluded from
the analysis. The number of galaxies used in the analysis of rotation
is seven, therefore.

\begin{table}
\caption{Table of line widths of LBGs measured from rest-frame optical
emission lines}
\label{table:resultsLBG2}
\centering
\begin{scriptsize}
\begin{tabular}{llllll} \\ \hline
Galaxy & $z$ & ${\cal R }$ & $\sigma$ & line & ref. \\
       &     &             &   (\kms) &      &      \\
  (1)  & (2) & (3)         & (4)      & (5)  & (6)  \\ \hline

CDFa D18           & 3.1122 & 23.74 & $79\pm7$   & \OIII & P01 \\
CDFa C8            & 3.0752 & 23.72 & $106\pm6$  & \OIII & P01 \\
CDFa C1            & 3.1147 & 23.53 & $\leqslant63$      & \OIII & P01 \\
Q0201$+$113 C6     & 3.0548 & 23.92 & $64\pm 4$  & \OIII & P01 \\
Q0256$-$000 C17    & 3.2796 & 23.89 & $53\pm4$   & \OIII & P01 \\
Q0347-383 C5       & 3.2337 & 23.82 & $69\pm4$   & \OIII & P01 \\
B2 0902$+$343 C12  & 3.3866 & 23.63 & $87\pm12$  & \OIII & P01 \\
West MMD11         & 2.9816 & 24.05 & $53\pm 5$  & \OIII & P01 \\
Q1422$+$231 D81    & 3.1037 & 23.41 & $116\pm8$  & \OIII & P01 \\
3C324 C3           & 3.2876 & 24.14 & $76\pm18$  & \OIII & P01 \\
SSA22a MD46        & 3.0855 & 23.30 & $67\pm6$   & \OIII & P01 \\
SSA22a D3          & 3.0687 & 23.37 & $113\pm7$  & \OIII & P01 \\
DSF 2237$+$116a C2 & 3.3176 & 23.55 & $100\pm4$  & \OIII & P01 \\
Q0000$-$263 D6     & 2.966  & 22.88 & $60\pm10$  & \OIII & P01 \\
B2 0902$+$343 C6   & 3.091  & 24.13 & $55\pm15$  & \OIII & P01 \\
MS1512-cB58        & 2.7290 & 24.10 & $81$       & \OIII & P01 \\

CDFb-BN88          & 2.2615 & 23.14 & $96\pm46$  & \ha   & E03 \\
Q0201-B13          & 2.1663 & 23.34 & $62\pm29$  & \ha   & E03 \\
Westphal-BX600     & 2.1607 & 23.94 & $181\pm24$ & \ha   & E03 \\
Q1623-BX376a       & 2.4085 & \multirow{2}{*}{23.31$^{\ast}$} &$261\pm72$ & \ha   & E03 \\
Q1623-BX376b       & 2.4085 &       & $<224$     & \ha   & E03 \\
Q1623-BX432        & 2.1817 & 24.58 & $51\pm22$  & \ha   & E03 \\
Q1623-BX447        & 2.1481 & 24.48 & $174\pm18$ & \ha   & E03 \\
Q1623-BX449        & 2.4188 & 24.86 & $141\pm94$ & \ha   & E03 \\
Q1623-BX511        & 2.2421 & 25.37 & $152\pm47$ & \ha   & E03 \\
Q1623-BX522        & 2.4757 & 24.50 & $<44$      & \ha   & E03 \\
Q1623-MD107        & 2.5373 & 25.35 & $<42$      & \ha   & E03 \\
Q1700-BX691        & 2.1895 & 25.33 & $170\pm18$ & \ha   & E03 \\
Q1700-BX717        & 2.4353 & 24.78 & $<60$      & \ha   & E03 \\
Q1700-MD103        & 2.3148 & 24.23 & $75\pm21$  & \ha   & E03 \\
Q1700-MD109        & 2.2942 & 25.46 & $87\pm35$  & \ha   & E03 \\
SSA22a-MD41        & 2.1713 & 23.31 & $107\pm15$ & \ha   & E03 \\
\hline
\end{tabular}
\begin{minipage}{170mm}

$^{\ast}$ Q1623-BX376 appears as a single extended object for photometry \\
but is resolved into two lines at the same redshift in the \ha\ spectroscopy \\
(see E03 for further details).

\end{minipage}
\end{scriptsize}
\medskip
\end{table}

\begin{table*}
\caption{Summary of rotation velocity data for LBGs}
\label{table:resultsLBG}
\centering
\begin{scriptsize}
\begin{tabular}{lllllllllll} \\ \hline
Galaxy & $z$ & ${\cal R}$ & $\sigma$  & $V_{\rm r}$ & $V_r/\sigma$ & $\theta$ & $r$ & slit width  &
 line & ref. \\
       &     &            &   (\kms)  & (\kms)      & & (arcsec) & (kpc) & 
(arcsec) & & \\
 (1) & (2) & (3) & (4) & (5) & (6) & (7) & (8) & (9) & (10) & (11) \\ \hline

Q0347-383 C5        & 3.2337 & 23.82 &$69\pm4$   & $31$  & 0.45 & 0.60 & 4.5 & 1.0 &\OIII & P01 \\
								 	      
SSA22a MD46         & 3.0855 & 23.30 &$67\pm6$   & $40$  & 0.60 & 0.59 & 4.5 & 1.0 &\OIII & P01 \\
								 	      
Westphal-BX600      & 2.1607 & 23.94 &$181\pm24$ & $210$ & 1.16 & 0.71 & 5.9 & 0.76&\ha   & E03 \\
								 	      
Q1623-BX447         & 2.1481 & 24.48 &$174\pm18$ & $160$ & 0.92 & 0.61 & 5.1 & 0.76&\ha   & E03 \\
								 	      
Q1623-BX511         & 2.2421 & 25.37 &$152\pm47$ & $80$  & 0.53 & 0.38 & 3.1 & 0.76&\ha   & E03 \\
								 	      
Q1700-BX691         & 2.1895 & 25.33 &$170\pm18$ & $220$ & 1.29 & 0.76 & 6.3 & 0.76&\ha   & E03 \\
								 	      
Q1700-MD103         & 2.3148 & 24.23 &$75\pm21$  & $100$ & 1.33 & 0.84 & 6.9 & 0.76&\ha   & E03 \\
								 	      
SSA22a-MD41$^{\ast}$& 2.1713 & 23.31 &$107\pm15$ & $150$ & 1.40 & 0.98 & 8.1 & 1.0 &\ha   & E03 \\

\hline
\end{tabular}
\begin{minipage}{170mm}
$^{\ast}$ SSA22a-MD41 is not used in the statistical analysis of rotation
curves as the slit was intentionally aligned along the galaxy major
axis.
\end{minipage}
\end{scriptsize}
\medskip
\end{table*}


\section{Analysis of rotation curves}

In this section we analyse the kinematic data for this small subsample
of seven LBGs. The aim is to investigate the relation between the two
quantities $\sigma$ and $V_r$. Rotation velocities are more difficult
to measure, requiring detection of emission lines in the fainter outer
parts of galaxies. Also, rotation velocities measure only the
projection of the useful quantity, the circular velocity $V_c$. Since
the inclination angles of LBGs cannot be measured reliably, data on
rotation velocities require a statistical analysis. The central line
widths, on the other hand, are easier to measure, and, to the extent
that they reflect random motions rather than rotation, provide the
kinematic quantity of interest (i.e. the one-dimensional velocity
dispersion) directly. The disadvantage, as stated earlier, is that the
measurements are confined to small radii. Therefore the purpose of our
analysis is to see whether the line width can be used as a predictor
of the circular velocity.

What do the central line widths of LBGs represent? The results of a
similar analysis of the kinematics of nearby starburst galaxies, by
Lehnert \& Heckman (1996), bear comparison. These authors plotted
[NII] nuclear line width against galaxy circular velocity, and found
no correlation. However it is not clear how relevant this analysis is
to the measurements of LBGs. Lehnert \& Heckman suggested that the
lack of correlation was because their nuclear measurements, scale 1
arcsec at $z\sim0.05$, $\equiv0.1$kpc, sample the steeply rising part
of the rotation curve. In contrast the line widths of LBGs sample
larger radii, scale 0.2 arcsec at $z\sim3$, $\equiv1.5$kpc, and
characterise the motion of the majority of the luminous matter in the
galaxies. All the same, the central line widths may not reflect random
motions, but could include a contribution from rotation. Rather than
attempt to interpret exactly what the central line widths measure, or,
indeed, to relate the results at high redshift to the results at low
redshift, we simply ask the question whether the central line widths
are useful.

The answer to this question is provided by Fig. 1, where we plot, as
the stepped solid line, the cumulative probability distribution
function (PDF) of the ratio $V_r/\sigma$ for the seven galaxies with
measured rotation velocities. Testing the MMW picture hinges on a
careful interpretation of this figure. To aid the interpretation we
have computed the predicted cumulative PDFs for some simple models. In
these models we suppose galaxies may be approximated as stellar disks
of negligible mass embedded in dark matter singular isothermal
spheres (SIS), characterised by the one-dimensional velocity dispersion,
$\sigma_{1D}$. The SIS has a flat rotation curve, and the circular
velocity is related to the one-dimensional velocity dispersion by
$V_c=\sqrt{2}\sigma_{1D}$. The dashed curves (1) and (2) in Fig. 1 are
the predicted cumulative PDFs of the ratio $V_r/\sigma_{1D}$ for the limiting 
cases (1) where the galaxy is much larger than the slit width (the
``large-galaxy case''), and (2) where the galaxy is smaller than the
slit width (the ``small-galaxy case'').  For the small-galaxy case,
the curve is shifted to higher values of $V_r/\sigma_{1D}$ relative to
the large-galaxy case. This is because, in the small-galaxy case, the
emission from material along the galaxy major axis, which has the
highest rotation velocity, is always recorded, whereas in the
large-galaxy case this only occurs when the slit is fortuitously
aligned along the major axis.

Before attempting to interpret these results we quantify the goodness
of fit between the data and the models, with the Kolmogorov-Smirnov
(KS) test. For the large-galaxy case we find a KS probability
$P(KS)=0.01$, and for the small-galaxy case we find $P(KS)=0.36$. The
dotted curve (3), in Fig. 1, is an intermediate case, where, for each
value of $V_r/\sigma_{1D}$, we have averaged the cumulative
probabilities for cases (1) and (2). For curve (3) we find
$P(KS)=0.42$. Curves (2) and (3) are therefore consistent with the
data, while curve (1) is inconsistent with the data.

To interpret the stepped line, we begin with the hypothesis that the
circular velocity in any galaxy may be predicted using the relation
$V_c=\sqrt{2}\sigma$. [Note, however, that, at this stage, we make no
attempt to connect the quantity $\sigma$ (measured from the LBG line
widths) and the quantity $\sigma_{1D}$.] Under this hypothesis, the
predicted cumulative PDF for $V_r/\sigma$ depends on the size of the
galaxies relative to the slit, and would lie between curves (1) and
(2). Since the projected angular diameters of the galaxies (average
value $2\overline{\theta}=1.3\arcsec$, Table 2) are comparable to, but
a little larger than, the slit width, if the hypothesis
$V_c=\sqrt{2}\sigma$ is a good one, we would expect the data to lie
closer to curve (2) than to curve (1), and therefore somewhere between
curves (3) and (2). Since the data do indeed fall between curves (3)
and (2), and since the KS probabilities for these two curves are both
satisfactory, we conclude that the data are consistent with the
hypothesis $V_c=\sqrt{2}\sigma$.

Since this conclusion is based on a small sample, to what extent could
this relation underestimate the galaxy circular velocities? We will
establish this by stretching the intermediate curve by a factor $F$.
In other words if we apply the relation $V_c=F\sqrt{2}\sigma$, how
large can $F$ be before the intermediate curve no longer matches the
data? Because the circular velocities predicted by MMW are large, in the
context of ruling out the MMW model this is a conservative assumption,
since we should really stretch a curve that lies between (3) and (2),
in which case the factor $F$ would be smaller. We find that when
$F=1.55$ the probability P(KS) falls to 0.05, which is to say that the
relation $V_c=\sqrt{2}\sigma$ underestimates the circular velocities
by 55 percent, at the most. This stretched curve is plotted as curve 4 in
Fig. 1.

Turning to the physical meaning of these results, one possible
interpretation is that the mass distribution in the central few kpc of
LBGs is approximately isothermal i.e. their rotation curves are
approximately flat. While this is perfectly plausible, this
interpretation relies on the identification of $\sigma$ with
$\sigma_{1D}$, in other words that the line widths reflect random
motions as opposed to rotation (or indeed any non-gravitational
motions). Higher resolution spectroscopic data, from adaptive-optics
fed spectrographs, will be required to confirm this conjecture. For
our purposes, however, it is sufficient to note simply that a good fit
to the distribution of rotation velocities for these seven galaxies is
obtained by applying the relation $V_c=\sqrt{2}\sigma$, which is to
say that, for a sample of LBGs, the distribution of central line widths 
is a good predictor of the distribution of the circular velocities at radii
of several kpc. This may or may not be true for individual galaxies.

\begin{figure}
\caption{Cumulative PDFs of the ratio $V_r/\sigma$ for the seven
galaxies from Table 2 (excluding SSA22a-MD41) plotted solid, and for
four models. Dashed: (1) SIS large-galaxy case, (2) SIS small-galaxy
case. Dotted: (3) SIS intermediate case (average of lines 1 and 2),
(4) intermediate case stretched by a factor $F=1.55$.}
\label{fig:cumf_normal}
\includegraphics[angle=90, width=1\columnwidth]{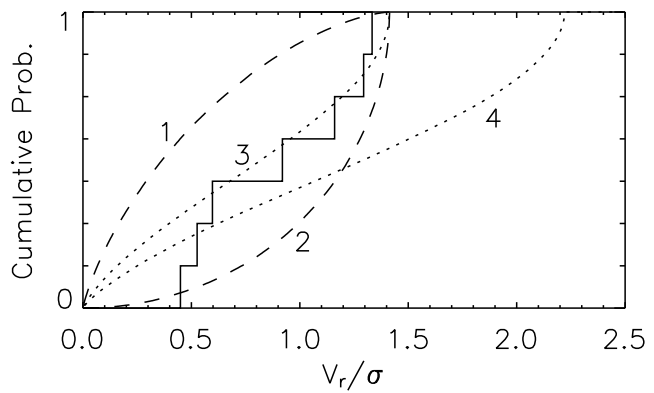}
\end{figure}

The galaxies of Table 2 are the subsample of galaxies in Table 1 for
which it was possible to measure rotation i.e for which ionised gas is
detectable a few kpc from the galaxy centre. For the present we assume this
subsample provides an unbiased estimate of the quantity
$V_c/\sigma$, and we will apply the relation $V_c=\sqrt{2}\sigma$ to
all the galaxies of Table 1. Possible biases are discussed in Section 4. Since the quantity $V_c/\sigma$ is a
dimensionless ratio it is unlikely to be a strong function of luminosity or redshift, so
even if the galaxies in the subsample of Table 2 differed
substantially from the galaxies of Table 1 in terms of these quantities, one might 
reasonably still apply the relation. In fact the
two samples are quite well matched. For the seven galaxies with
rotation measures the average quantities are
${\cal\overline{R}}=24.4$, and $\overline{z}=2.48$, and for the
galaxies in Table 1 the average quantities are
${\cal\overline{R}}=24.1$, and $\overline{z}=2.71$.

The average projected radius at which the rotation velocity was
measured for the seven galaxies used in the analysis of this section
(Table 2) is $\overline{r}=5.2$ kpc. Accounting for projection effects,
this corresponds to an average deprojected radius of
$\overline{R}\sim7$ kpc. In the following section we compare the
distribution of circular velocities at a radius of 7 kpc predicted by
MMW, for the dark matter, against the distribution of circular
velocities at that radius inferred for the galaxies of Table 1, by
applying the relation $V_c=\sqrt{2}\sigma$.


\section{Comparison against theoretical predictions}

In this section we first summarise the predictions of MMW of the
halo circular velocities of LBGs, and then compare these predictions
against the data of Table 1. Specifically the MMW predictions are for
LBGs at $z=3$, brighter than ${\cal R}=25.5$. Using the luminosity
function provided by Adelberger and Steidel (2000), the average
magnitude of this population is $\overline{{\cal R}}=24.9$. The
average quantities for the observed galaxies in Table 1 are
$\overline{z}=2.71$ and ${\cal\overline{R}}=24.1$. Therefore the
average redshift of the observed galaxies is quite close to the
comparison datum $z=3$, while the average apparent magnitude is
brighter by 0.8 mag. One might expect a sample of galaxies with
average apparent magnitude $\overline{{\cal R}}=24.9$ to have smaller
circular velocities on average than the sample of Table 1. All the
same P01 found no correlation between central line width and galaxy
luminosity in their sample.

MMW begin with the assumption that LBGs are the central galaxies of
the most massive dark-matter halos. The Press-Schechter formalism
yields the mass function, and the density profiles are assumed to be
of NFW form (Navarro et al., 1997). In outline, matching to
the observed volume density of LBGs brighter than ${\cal R}=25.5$,
provided by Adelberger et al. (1998), specifies the distribution of
halo circular velocities\footnote{The actual model is more detailed
than implied here.}, which is plotted in fig. 4 of MMW (the solid line
plots the results for the $\Lambda$ cosmology adopted in this
paper). Here the halo circular velocity $V_{h}$ is the value at the
virial radius $R_{h}$, which is the radius at which the average density of
the halo inside that radius is equal to 200 times the critical density
at that epoch, $\rho_c=3H(z)^2/(8\pi G)$. Therefore $V_{h}$ and
$R_{h}$ are simply related by $V_{h}=10H(z)R_h$. Note that the halo
mass $M_{H}=V_{h}^2R_h/G$, is proportional to $V_{h}^3$.

For the NFW profile, the circular velocity $V_c$ at any radius,
specified by $x=R/R_h$, is related to $V_h$ by:

\[ 
\Big(\frac{V_c}{V_h}\Big)^2=\frac{1}{x}\frac{\ln(1+cx)-(cx)/(1+cx)}
{\ln(1+c)-c/(1+c)}.
\]

Here the parameter $c$, called the `concentration', is the ratio
$R_h/R_s$, where $R_s$ is the scale radius of the NFW profile. The
typical predicted halo circular velocities are in the range
$200-400$\kms. Based on the results of Zhao et al. (2003), a suitable
value for the concentration at $z\sim3$, for these values of $V_h$, is
$c=4$ (Mo, priv. comm.). Using this value we can now compute the ratio
$V_c/V_h$ at the radius of interest $R=7$ kpc, for different values of
$V_h$, and transform the distribution of $V_h$ to the distribution of
$V_c$. For example, for halo circular velocities of 200 and 400 \kms,
we find $V_c/V_h=0.81$ and 0.64, respectively.

In Fig. 2 we compare the predicted circular velocities of LBGs against
the measured values. In this plot line (1) shows the cumulative
distribution of $V_h$, predicted by MMW, and line (2) shows the
cumulative distribution of $V_c$ at $R=7$ kpc, computed as described
above. Line (3) is the cumulative distribution of measured circular
velocities of LBGs at $R=7$ kpc, produced by applying the relation
$V_c=\sqrt{2}\sigma$ to the data of Table 1. The measured circular
velocities are much smaller than the predicted values. The probability that
the two distributions are drawn from the same population is negligible. Allowing
for appropriate scatter in the parameter $c$ does not alter this conclusion. 
We have computed the scaling by which the circular
velocities of line (2) should be divided, to provide the best fit to
line (3), as measured by the KS test. This factor, which we call
$Q$, is 1.8. We can associate an indicative $1\sigma$ uncertainty of
0.4 with this factor, by recalling the results of Section 3, where we
found that, at 5 percent confidence (i.e. $1.65\sigma$), the relation
$V_c=\sqrt{2}\sigma$ could underestimate the true circular velocities
by a factor as large as $F=1.55$. To summarise, we find that the
circular velocities at a radius $R=7$ kpc, predicted by MMW, are a
factor $Q=1.8\pm0.4$ larger than the measured circular velocities of
LBGs at $z\sim 3$. This rather large uncertainty in $Q$ is a
consequence of: i) the small number of galaxies with rotation
measures, ii) the fact that the slit was not aligned along the major
axis, but fell at a random angle.

\begin{figure}
\caption{Measured and predicted cumulative PDFs of the circular
velocity of LBGs at $z=3$: Line (1) MMW prediction for $V_h$, line (2)
MMW prediction for $V_c$ at $R=7$ kpc, line (3) measured values of
$V_c$ at $R=7$ kpc for the 32 galaxies of Table 1, derived using the
relation $V_c=\sqrt{2}\sigma$, line (4) SPF
collisional starburst prediction for $V_c$.}
\label{fig:charming}
\includegraphics[angle=90, width=1\columnwidth]{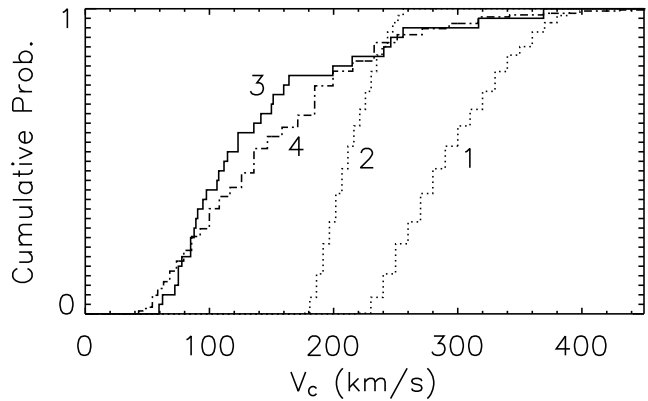}
\end{figure}

Could this discrepancy between the predicted and measured circular
velocities of LBGs be explained by a bias in the data? E03 note one
such effect. For the small-galaxy case, at any particular projected
radius, the line flux includes contributions both from the point on
the galaxy major axis (where the rotation velocity is a maximum), and
from points off the major axis (where the rotation velocities are
smaller).  For this reason some of the rotation velocities measured by
Erb may be smaller than the true values, and this would reduce the
discrepancy between theory and observation. This effect is mitigated
to some extent by the falling surface brightness profiles of galaxies,
since at any projected radius the surface brightness is then highest
along the major axis. Another effect which we have not considered acts
in the opposite sense, and widens the discrepancy between theory and
observation. For galaxies larger than the slit width, the projected
radius of the flux in the slit is largest when the slit is aligned
along the major axis. These are also the galaxies (for the given
inclination angle) for which the measured rotation velocity is
maximised. For galaxies oriented away from the slit, the projected
radius will be smaller, and then the elongation along the slit may not
be detectable. These galaxies have smaller rotation velocities, and
may be missing from the sample. Indeed it is noticeable that there are
no LBGs in Table 2 for which $V_r/\sigma<0.4$. To test this we made a simple simulation of 
galaxies of radius  $\sim1$ arcsec, with random inclination angles, and
observed in $0.5$ arcsec seeing, at random slit positions. We discarded objects with projected 
size along the slit smaller than the seeing. The simulated curve closely follows the stepped line in Figure 1, 
the limit on the projected size causing the dearth of objects with small rotation velocities. This indicates  
that the apparent deficit visible in Fig. 1 may be a
selection effect. If so the factor $F$ computed above (and so the
uncertainty on $Q$ also) will have been overestimated.

There are number of other reasons to believe that the discrepancy 
between theory and observation, may be more significant than estimated
above. These include:

\begin{enumerate}
\item The effect of the baryons has not been included in the
prediction. Because of dissipation, baryons sink toward the centres of
the dark matter halos, deepening the potential well, and raising the
circular velocities above the values for the dark matter only.
\item In the data of Table 1, where an upper limit to the central line
width has been quoted, we have computed the circular velocity for this
value, whereas the true circular velocity will be smaller.
\item The MMW circular velocities were computed for a population with
an average apparent magnitude 0.8 mag. fainter than the average apparent
magnitude of the galaxies in Table 1.
\item As explained in Section 3, we were conservative in computing the
factor $F$, since to compute $F$ we stretched curve (3) of Fig. 1. A
curve between curves (3) and (2) would have produced a smaller value
of $F$, and therefore a smaller value for the uncertainty of $Q$,
increasing the discrepancy between theory and observation.
\end{enumerate}

Taking these factors into consideration, this analysis indicates that
the simple model of LBGs marking the centres of the most massive dark
matter halos is probably incorrect.

We now consider the alternative picture of LBGs as galaxies brightened
by starbursts, triggered by collisions. 
We take the predicted distribution of halo masses for the SPF
collisional starburst model, from Primack et al. (2003), and convert
each mass to a circular velocity using equations from the appendix of
Somerville and Primack (1999), for the truncated isothermal sphere
used in their models. Note that this model has a flat rotation curve 
and their definition of virial radius differs slightly from that of MMW. 
The distribution of predicted circular
velocities is plotted as line (4) in Fig. 2. The model provides a good fit 
to the measurements, with associated probability P(KS)=0.26. We also computed
the circular velocities at R=7kpc assuming an NFW profile, and found values about 10 percent smaller. 

It has proved difficult to discriminate between the two models for
LBGs at $z\sim3$ discussed in this paper, on the basis of their
luminosities and clustering properties (Wechsler et al., 2001). The
evidence assembled here from the kinematics clearly favours the
collisional starburst picture. Our analysis has relied on the measurement of
rotation velocities for just seven galaxies. Spectroscopic
measurements, with high spatial resolution, along the galaxy major
axis, for a larger sample, will be useful in further clarifying the
nature of LBGs.


\section*{Acknowledgments}

We thank Rachel Somerville and Houjun Mo for useful communications 
regarding the mass distribution of various LBG models. SJW[1] thanks 
Oliver Warren for his patient mathematical guidance and Thomas 
Babbedge and Alexander King for their helpful comments on the text.


\bsp

\label{lastpage}


\begin{thebibliography}{}

\bibitem[\protect\citeauthoryear{Adelberger et al. }{1998}]{A98}
Adelberger K. L., Steidel C. C., Giavalisco M., Dickinson M. E., 
Pettini M., Kellogg M., 1998, ApJ, 505, 18

\bibitem[\protect\citeauthoryear{Adelberger \& Steidel}{2000}]{A00}
Adelberger K. L., Steidel C. C., 2000, ApJ, 544, 218

\bibitem[\protect\citeauthoryear{Babcock}{1939}]{BA39} 
Babcock H. W., 1939, Lick Obs. Bull., 19, 41

\bibitem[\protect\citeauthoryear{de Blok, Bosma \& McGaugh}{2003}]{DB03} 
de Blok W. J. G., Bosma A., McGaugh S., 2003, MNRAS, 340, 657

\bibitem[\protect\citeauthoryear{Erb et al. }{2003}]{E03} 
Erb D. K., Shapley A. E., Steidel C. C., Pettini M., Adelberger K. L., Hunt
M. P., Moorwood A. F. M., Cuby J.-G., 2003, ApJ, 591, 110

\bibitem[\protect\citeauthoryear{Giavalisco et al. }{1998}]{GS98}
Giavalisco M., Steidel C. C., Adelberger K. L., Dickinson M. E.,
Pettini M., Kellogg M., 1998, ApJ, 503, 543

\bibitem[\protect\citeauthoryear{Kolatt et al. }{1999}]{K99} 
Kolatt T. S. et al., 1999, ApJ, 523, L109

\bibitem[\protect\citeauthoryear{Lacey \& Cole}{1993}]{LC93} 
Lacey C., Cole S., 1993, MNRAS, 262, 627

\bibitem[\protect\citeauthoryear{Lehnert \& Heckman}{1996}]{MS96}
Lehnert M. D., Heckman T. M., 1996, ApJ, 472, 546

\bibitem[\protect\citeauthoryear{Marleau \& Simard}{1998}]{MS98}
Marleau F. R., Simard L., 1998, ApJ, 507, 585

\bibitem[\protect\citeauthoryear{Mo, Mao \& White }{1999}]{MMW99} 
Mo H. J., Mao S., White S. D. M., 1999, MNRAS, 304, 175

\bibitem[\protect\citeauthoryear{Navarro, Frenk \& White }{1997}]{NFW97}
Navarro J. F., Frenk C. S., White S. D. M., 1997, ApJ, 490, 493

\bibitem[\protect\citeauthoryear{Pettini et al. }{2001}]{P01} 
Pettini M., Shapley A. E., Steidel C. C., Cuby J.-G., Dickinson M., Moorwood
A. F. M., Adelberger K. L., Giavalisco M., 2001, ApJ, 554, 981

\bibitem[\protect\citeauthoryear{Primack, Wechsler \& Somerville
}{2003}]{PWS03} Primack J. R., Wechsler R. H., Somerville R. S., 2003,
in Bender R. \& Renzini A., eds. The Mass of Galaxies at Low and High
Redshift. Proceedings of the ESO Workshop held in Venice, October 2001

\bibitem[\protect\citeauthoryear{Rubin, Thonnard \& Ford}{1978}]{RTF78} 
Rubin V. C., Thonnard N., Ford W. K. Jr., 1978, ApJ, 225, L107

\bibitem[\protect\citeauthoryear{Somerville, \& Primack}{1999}]{SP99}
Somerville R. S., Primack J. R., 1999, MNRAS, 310, 1087

\bibitem[\protect\citeauthoryear{Somerville, Primack \& Faber
}{2001}]{SPF01} Somerville R. S., Primack J. R., Faber S. M., 2001,
MNRAS, 320, 504

\bibitem[\protect\citeauthoryear{Steidel \& Hamilton }{1993}]{SH1993}
Steidel C. C., Hamilton D., 1993, AJ, 105, 2017

\bibitem[\protect\citeauthoryear{Steidel et al.}{1996}]{ST1996}
Steidel C. C., Giavalisco M., Pettini M., Dickinson M., Adelberger
K. L., 1996, ApJ, 462, L17

\bibitem[\protect\citeauthoryear{Steidel et al.}{1998}]{ST1998}
Steidel C. C., Adelberger K. L., Dickinson M., Giavalisco M., Pettini
M., Kellogg M., 1998, ApJ, 492, 428

\bibitem[\protect\citeauthoryear{van Albada et al.
}{1985}]{VA85} van Albada T. S., Bahcall J. N., Begeman K., Sancisi
R., 1985, ApJ, 295, 305

\bibitem[\protect\citeauthoryear{Wechsler et al. }{2001}]{W01}
Wechsler R. H., Somerville R. S., Bullock J. S., Kolatt T. S., Primack
J. R., Blumenthal G. R., Dekel A., 2001, ApJ, 554, 85

\bibitem[\protect\citeauthoryear{Zhao et al.}{2003}]{Z03}
Zhao D. H., Mo H. J., Jing Y. P., B\"{o}rner G., 2003, MNRAS, 339, 12

\end{thebibliography}
\end{document}